\documentstyle[12pt,graphicx]{article}

\begin{document}


\title{ Double-logarithmic (Sudakov) asymptotics at the theory of 
electroweak interactions 
\footnote{talk given by B.I.~Ermolaev} }


\author{B.I.~Ermolaev$^{ab}$, M.~Greco$^c$, S.M.~Oliveira$^b$ and
  S.I.~Troyan$^d$\vspace*{0.2 cm} \\ 
\small{$^a$Ioffe Physico-Technical Institute, 194021
 St.Petersburg, Russia}\\
\small{$^b$CFTC, University of Lisbon
Av. Prof. Gama Pinto 2, P-1649-003 Lisbon, Portugal}\\
\small{$^c$University Rome-3, Rome, Italy}\\
\small{$^d$St.Petersburg Institute of 
Nuclear Physics, 188300 Gatchina, Russia}}

\maketitle

\begin{abstract}
Accounting for double-logarithmic contributions to 
high-energy ($\gg 100$ GeV) $e^+ e^-$ annihilation 
into a quark  or a lepton pair in 
the kinematics where the final particles are colinear to the $e^+e^-$
beams leading to a sizable difference 
between the forward and backward scattering amplitudes, i.e. to the 
forward-backward asymmetry. When the annihilation is accompanied by emission 
of $n$ electroweak bosons in the multi-Regge kinematics, it turns out that 
the cross sections of the photon and $Z$ production have the identical energy 
dependence and asymptotically 
their ratio depends only on the Weinberg angle 
(is equal to $\tan^{2n} \theta_W$) 
whereas the energy dependence of 
the cross section of the $W$ production is suppressed 
by factor $s^{-0.4}$ compared to them.   
\end{abstract}

\section{Introduction}

The double-logarithmic approximation (DLA) was introduced into the particle 
physics by V.V.~Sudakov who first found that the most important  
radiative corrections to the form factor $f(q^2)$ of electron at large 
$q^2$ are the 
double-logarithmic (DL) ones, i.e. 
$\sim (\alpha \ln^2(q^2/ m^2))^n$~ ($n =1,2,..$). 
with $m$ being a mass scale,. 
After accounting them to all orders in $\alpha$, it turns out\cite{sud} that 
asymptotically 
\begin{equation}
\label{sud}
f(q^2) \sim e^{-(\alpha/ 4\pi) \ln^2(q^2/m^²)}
\end{equation}   
when $q^2 \gg m^2$. The next important step towards studying DL asymptotics in 
QED was done in Refs.~\cite{ggfl}. After that, calculating in DLA has  
become rather technology than art. Studying QCD scattering amplitudes showed 
that there is no big technical 
difference between the QED and QCD as for calculating 
amplitudes of elastic processes (see e.g. Ref.~\cite{kl}) 
whereas inelastic (radiative) QCD -amplitudes 
are much harder to calculate (see e.g. Refs.~\cite{kf}). The methods of 
calculating the DL asymptotics can be 
applied also to electroweak (EW) processes providing the total energy is high 
enough to neglect masses of the electroweak bosons. At such huge energies 
($\gg 100$ GeV),  
many important technical details learnt from QED and QCD can be used for 
calculating EW amplitudes\cite{flmm}. In the present talk I discuss  
DL asymptotics for  $e^+ e^-$ annihilation 
into a quark-anitquark  or a lepton-antilepton 
pair (the elastic annihilation)\cite{egt} and the inelastic 
annihilation\cite{eot} where $e^+ e^-$ annihilate into a quark-antiquark 
(lepton-antilepton) pair and 
electroweak bosons. 

\section{DL contributions to elastic $e^+ e^-$ -annihilation 
into quarks and leptons}

The conventional way for considering  $e^+ e^-$ -annihilation into 
$\to q \bar{q}$ consists of two steps: the first one is the 
assumption that this process is mediated by a single 
virtual photon exchange: $e^+ e^- \to \gamma^* \to q \bar{q}$; the 
second step is calculating QCD radiative corrections. However, 
electroweak radiative corrections can also be sizable when this 
process is considered in some 
particular kinematic regions. These are the forward and backward   
kinematics. If the scattering angle\footnote{Through this paper 
when we refer to angles, we imply the angles in cmf.} 
between momenta of the 
initial electron (positron) and of the final particles with 
the negative (positive) electric 
charges is $\ll 1$, it is the forward kinematics. 
The case when this angle is $\sim \pi$ is the backward kinematics. 
Both these 
kinematics are of the Regge type and effect of accounting for the DL 
radiative corrections to all orders in the electroweak couplings 
in these kinematics can be interpreted as exchanges 
with Reggeons propagating in the cross channel. It means that 
expressions for the forward and backward scattering amplitudes  can 
be represented in the form of the Sommerfeld-Wotson (SW) integral:
\begin{equation}
\label{sw}
M^{(\pm)}_j(s/\mu^2) =
\int_{-\imath \infty}^{\imath \infty}
\frac{d\omega}{2\pi\imath} \left(\frac{s}{\mu^2}\right)^{\omega}\,
\xi^{(\pm)}(\omega)\,
F_j^{(\pm)}(\omega)
\end{equation}
where the signs $\pm$ refer to the signatures of the amplitudes, 
the signature factors 
$\xi^{(+)} \approx 1,~~ \xi^{(-)} \approx~ \imath\pi\omega/2$;  
$F_j^{(\pm)}(\omega)$ is called the SW  amplitude (the partial wave); 
the Mandelstam variable $s$ and the mass scale $\mu$ obey 
$\sqrt{s} \gg \mu \geq 100$GeV.  The integration contour in Eq.~(\ref{sw}) 
runs to the right from singularities of $F_j^{(\pm)}(\omega)$ . 
Subscript $j$ enumerates both the flavours of the produced quarks and 
the EW isospin state in the cross channel.  
For example, for the forward $e^+_R(p_2)e^-_L(p_1)$ annihilation into 
$u_L(p'_2) \bar{u}_R(p'_1)$, 
\begin{eqnarray}
\label{mu}
M^{(+)}(\rho,\eta) &=& a\,
\exp\left[-\,\frac{1}{8\pi^2} \,
\left(\frac32\, g^2 + \frac{Y_e^2+Y_q^2}{4}\, g'^2 \right)
\, \frac{\eta^2}{2}\right]  \nonumber \\
& &\times \int_{-\imath \infty}^{\imath \infty} \frac{dl}{2\pi\imath}\,
e^{\lambda l (\rho-\eta)}
\frac{D_{p-1}(l+\lambda\eta)}
{D_{p}(l+\lambda\eta)}
\end{eqnarray}
where $g$ and $g'$ are the couplings, $Y_e$ ~($Y_q$) is the hypercharge of 
electron (quark),~ $a = (-3g^2 + g'^2Y_eY_q)/4$,~ 
$\lambda = -g'(Y_e + Y_q)/2$,~ $p = -a/\lambda^2$,~ 
$\rho = \ln(s/\mu^2),~\eta = \ln(-t/\mu^2)$ and $D_p$ are 
the Parabolic cylinder functions. Eq.~(\ref{mu}) is obtained for the 
kinematic region  $s = (p_1 + p_2)^2 \gg -t = -(p_2 - p'_1)^2$.   

The exponent in Eq.~(\ref{mu}) is the electroweak Sudakov form factor for 
this process. It accumulates the softest radiative DL corrections, with 
virtualities of the virtual EW bosons $\leq -t$. The harder DL 
contributions are collected in the SW integral.     
Singularities of the integrand in  Eq.~(\ref{mu}) are the zeros of $D_p$. 
Forward scattering amplitudes for $e^+e^-$ annihilation into quarks of other 
chiralities and flavours are represented by similar expressions. 
The only difference is in 
the values of factors $a$, $p$, $\lambda$. The same is true for the 
backward scattering amplitudes.       

If $F_j^{(\pm)}(\omega)$ are singular when    
$\omega = \Delta_{j_{r}}^{(\pm)}, ~(r = 1,...)$, 
then asymptotic dependence of $M^{(\pm)}_j$ on
$s$ is
\begin{equation}
\label{as}
M^{(\pm)}_j(s/\mu^2) \sim \sum_r (s/\mu^2)^{\Delta_{j_{r}}^{(\pm)}} ~
\end{equation}  
and $\Delta_{j_{r}}^{(\pm)}$  are  the intercepts of the Reggeons.  
Ref.~\cite{egt} states that  
the value of the intercepts depends also on   
flavours and chiralities of the final quarks. 
It turns out\cite{egt} that all intercepts for the backward amplitudes 
are negative whereas a part of intercepts for the forward amplitudes 
are positive. Therefore, the backward amplitudes rapidly fall when $s$ 
increases whereas the forward amplitudes slowly grow with $s$. 
This result can be called as the forward-backward charge asymmetry.  
In particular, the largest intercepts of the 
forward positive signature amplitudes $M_j$ 
(we drop the superscript ``+'') are  
$\Delta_u = 0.11$ for $e^-_L e^+_R \to u_L \bar{u}_R$ and  
$\Delta_d = 0.08$ for $e^-_L e^+_R \to d_L \bar{d}_R$.  
The notation  
$q_L ~(q_R)$ means that the particle $q$ is left-handed (right-handed). 
The other intercepts are lesser. Defining the asymmetry factor A in 
terms of the forward and backward cross sections $d\sigma_{F,B}$ of 
detecting the quarks in the forward (backward) cones with very small opening 
angles $\theta <  M_Z/\sqrt{s}$ :  
\begin{equation}
\label{asym}
A = [d\sigma_F - d\sigma_B]/[d\sigma_F + d\sigma_B]
\end{equation}   
where $d\sigma_{F(B)}$ stands for forward (backward) differential cross 
section. Performing numerical calculations, we arrive at the result  
plotted in Fig.~1. The difference between the forward and backward 
scattering amplitudes leads also to the fact that the average electric 
charge of the produced 
hadrons in the cone around of the $e^-$ -beam ($e^+$ -beam) 
is negative (positive) and the value of the average charge 
grows with energy as shown in Fig.~2. 
It is possible to apply the plots of Figs.~1 and 2, to the situation 
when the produced quarks are in a wider angular region  
$1 \ll \theta <  M_Z/\sqrt{s}$. To this end one should replace $\sqrt{s}$ in 
these Figs. by $M_Z/ \theta$.       

\begin{figure}[htbp]
\begin{center}
\begin{picture}(260,160)
\includegraphics{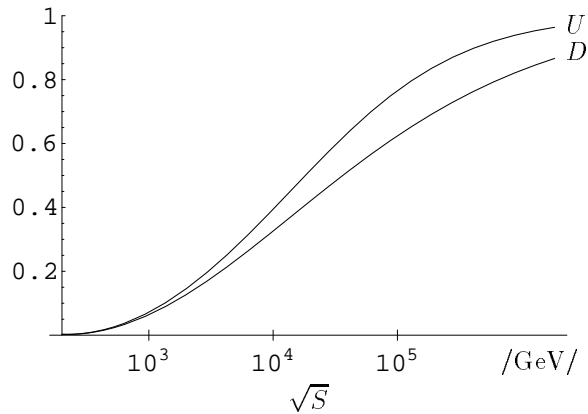}
\end{picture}
\end{center}
\caption{
Asymmetry $A$ for $e^+e^-$ annihilation into quarks of different flavours
in the ``collinear angular region'' in DLA.}
\label{fig3}
\end{figure}
\begin{figure}[htbp]
\begin{center}
\begin{picture}(290,205)
\includegraphics{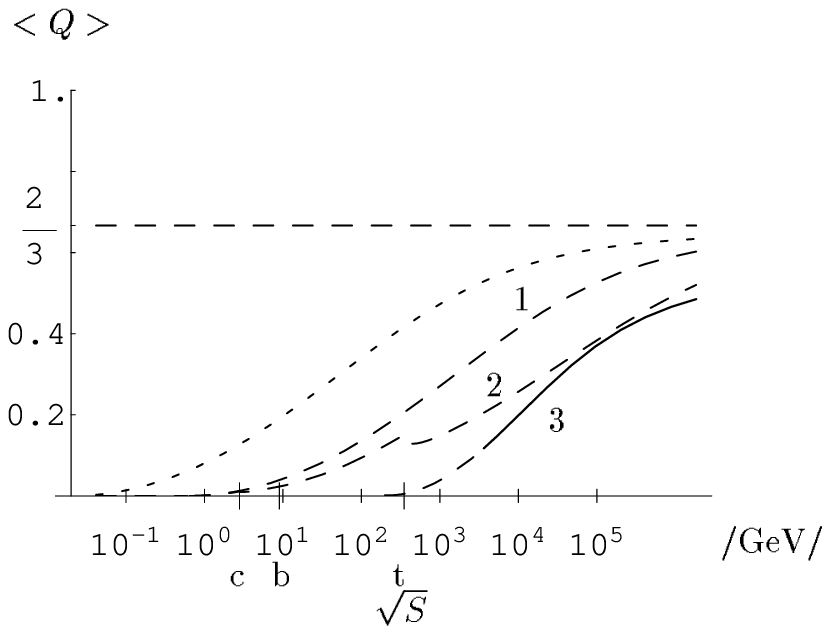}
\end{picture}
\end{center}
\caption{Average electric charge $<Q>$ of the hadron flow detected inside
a narrow cone $\theta< M/\sqrt{s}$ in the direction of
$e^+$~-beam.  Short-dashed curve corresponds to the case of multiphoton
annihilation to $u,d$~-current quarks with $M\equiv\mu=0.01$~Gev in QED.
Dashed curve 1 corresponds to $u,d$~-constituent quarks with
$M\equiv\mu=0.3$~GeV also in QED.
Curves 2 and 3 account for the all quark flavours
produced in $e^+e^-$~-annihilation: the curve 2 is calculated in
QED while the curve 3  corresponds to  all
EW -bosons exchanged in DLA with $M=M_Z$.
 Curve 2 shows how $<Q>$ would rise
without account of EW interactions. The dashed part of the curve
3 corresponds to the region where subleading corrections to DLA 
could be important.
  The dashed horizontal line shows the asymptotic value of
$<Q>$ as the $u$~-quark contribution is dominating.}
\label{fig4}
\end{figure}

\section{Inelastic $e^+e^-$ -annihilation into quarks}

When $e^+e^-$ annihilate into $q \bar{q}$ and electroweak bosons, with 
the final particles produced in the multi-Regge kinematics, there also 
appear DL electroweak corrections. 
The essence of the multi-Regge kinematics is that the longitudinal 
momenta of the produced particles are much greater than their 
transverse momenta. On the other hand, the transverse momenta 
$k_{i \perp}$ are assumed 
to be much greater than $M_{W,Z}$ so that all emission angles are $\ll 1$.     
With this assumption, the spontaneous broken $SU(2)U(1)$ symmetry 
in many respects can be 
regarded as restored. In particular, it becomes more convenient to 
consider emission of the isoscalar $A_0$ and isovector $A_{1,2,3}$ 
gauge fields and then to proceed to the $\gamma, W, Z$ emission, using 
the standard relations between these two sets. Also it 
makes possible to use arguments of 
Refs.~\cite{el2} where the multi-Regge amplitudes for gluon production 
were calculated. It turns out\cite{eot} that amplitudes for the $\gamma$ 
and $Z$  
production are governed by both the isoscalar and isovector Reggeons 
(with the intercepts $0.11$ and $0.08$)  
propagating in the cross channels, whereas the $W$ production is controlled 
by the isovector Reggeons only, with the smaller 
($-0.08$ and $-0.27$) intercepts. It means that the cross sections of 
the photon and the $Z$ production have identical energy dependence. 
The only difference between them is due to the different EW couplings, 
so that asymptotically (at energies $\sqrt{s} \geq 10^6$~Gev)
\begin{equation}
\label{gammaz}
\sigma^{nZ}(s)/\sigma^{n\gamma}(s) = \tan^{2n} \theta_W
\end{equation}     
whereas
\begin{equation}
\label{gammaz}
\sigma^{nW}(s)/\sigma^{n\gamma}(s) \sim s^{-0.4} ~.
\end{equation}  

Result of numerical calculations for these cross sections 
in the case of single boson production, covering the  
energy range from $10^3$ to $10^7$~GeV, are shown in Figs.~3 and 4. 

\begin{figure}[htbp]
\begin{center}
\begin{picture}(240,200)
\put(0,10){
\includegraphics{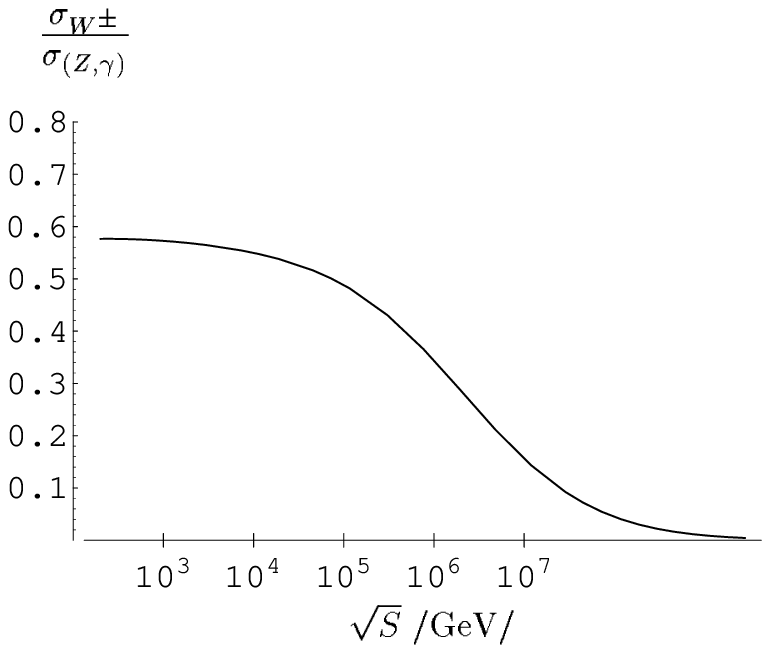}
}
\end{picture}
\end{center}
\caption{Total energy dependence of $W^{\pm}$ to
$(Z,\gamma)$ rate in $e^+e^-$ annihilation.}
\label{fig6}
\end{figure}

\begin{figure}[htbp]
\begin{center}
\begin{picture}(240,200)
\put(0,10){
\includegraphics{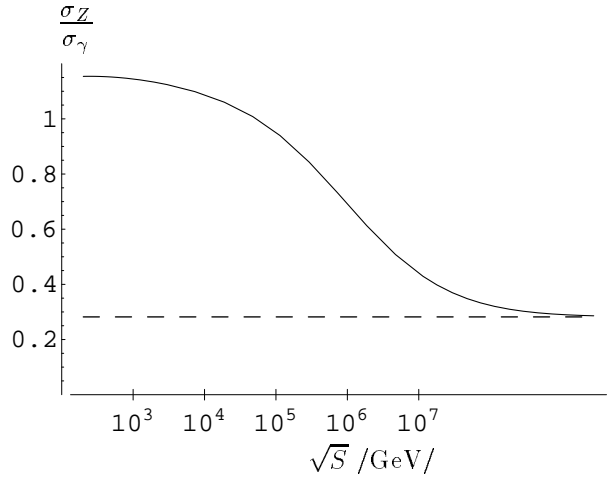}
}
\end{picture}
\end{center}
\caption{Total energy dependence of $Z$ to
$\gamma$ rate in $e^+e^-$ annihilation. The dashed line shows the
asymptotical value of the ratio: $\tan^2\theta_W\approx 0.28$~.}
\label{fig7}
\end{figure}

\section{Acknowledgement}
The work is supported by grants CERN/FIS/43652/2001, INTAS-97-30494,
SFRH/BD/6455/2001 and RFBR~00-15-96610.


\begin{thebibliography}{99}

\bibitem{sud}  V.V.~Sudakov. Sov. Phys. JETP 3(1956)65.  

\bibitem{ggfl} V.N.~Gribov, V.G.~Gorshkov, G.V.~Frolov, L.N.~Lipatov.
Sov.J.Nucl.Phys. 6(1968)95; ibid 6(1968)262.

\bibitem{kl} J.J.~Carazone, E.C.~Poggio and H.R.~Quinn. Phys. Rev. 
D 11(1975)2286; J.M.~Cornwell and G.~Tiktopolous. 
Phys. Rev. Lett. 35(1975)338; V.V.~Belokurov and N.I.~Usyukina. 
Phys. Lett. B94(1980)251; 
R.~Kirschner, L.N.~Lipatov. Nucl.Phys.B 213(1983)122.

\bibitem{kf} E.A.~Kuraev and V.S.~Fadin Yad. Fiz. 27(1978)1107;  
B.I.~Ermolaev and V.S.~Fadin JETP Letters 33(1981)269; 
\bibitem{efl} B.I.~Ermolaev, V.S.~Fadin, L.N.~Lipatov. Yad. Fiz. 45(1987)817.

\bibitem{flmm} V.S. Fadin, L.N. Lipatov, A.D.~Martin, and M.~Melles,
Phys.Rev. D{\bf61} (2000) 094002.

\bibitem{egt} B.I.~Ermolaev, M.~Greco, S.I.~Troyan. hep-ph/0205260.

\bibitem{eot} B.I.~Ermolaev, S.M.~Oliveira and S.I.~Troyan. 
hep-ph/0201159.  

\bibitem{el1} B.I.~Ermolaev, L.N.~Lipatov. Sov.J.Nucl.Phys. 47(1988)841.

\bibitem{el2} B.I.~Ermolaev, L.N.~Lipatov. Sov.J.Nucl.Phys. 48(1988)715;
Int.J.Mod.Phys.A4(1989)3147.

\end{thebibliography}
\end{document}